\documentclass[journal=jctc,manuscript=article]{achemso}

\usepackage{amsmath, amssymb}
\usepackage{mathtools}
\usepackage{mathrsfs}

\usepackage{xcolor}
\usepackage{float, graphicx}

\usepackage{dutchcal}

\newcommand{\mf}{\mathbf}
\newcommand{\lb}{\left(}
\newcommand{\rb}{\right)}
\newcommand{\lsq}{\left[}
\newcommand{\rsq}{\right]}

\author{Artem A. Finenko}
\affiliation{Department of Chemistry, Lomonosov Moscow State University, GSP-1, 1-3 Leninskiye Gory, Moscow 119991, Russia}
\alsoaffiliation{Institute of Quantum Physics, Irkutsk National Research Technical University, 83 Lermontov str., Irkutsk 664074, Russia}
\alsoaffiliation{Institute of Applied Physics of the Russian Academy of Sciences, 46 Ulyanov str., Nyzhny Novgorod 603950, Russia}
\email{artfin@mail.ru}

\title{Accurate neural-network-based fitting of full-dimensional two-body potential energy surfaces}
   
\begin{document}

\begin{abstract}
We describe the development of machine-learned potentials of atmospheric gases with flexible monomers for molecular simulations. A recently suggested permutationally invariant polynomial neural network (PIP-NN) approach is utilized to represent the full-dimensional two-body component of the dimer energy. To ensure the asymptotic zero-interaction limit, a tailored subset of the full invariant polynomial basis set is utilized and their variables are modified to achieve a better fit of the correct asymptotic behavior at a long range. The new technique is used to build full-dimensional potentials for the two-body N$_2-$Ar and N$_2-$CH$_4$ interactions by fitting databases of \textit{ab initio} energies calculated at the coupled-cluster level of theory. The second virial coefficients with full account of molecular flexibility effects are then calculated within the classical framework using the PIP-NN potential surfaces. 
To showcase the advantages of the PIP-NN method, we compare its accuracy and computational efficiency to several kernel-based and neural-network-based approaches using the MD17 database of energies and forces for ethanol. For large training set sizes, the PIP-NN models attain the best accuracy among examined models, and the computation time is shown to be comparable to that of the PIP regression model and several orders of magnitude faster than the quickest alternatives.
\end{abstract}

\section{Introduction}
\label{sec:introduction}

A potential energy surface (PES) is the key concept in computational chemistry that enables understanding the processes on an atomistic scale. The continuous changes within atomistic systems are often thought of as being driven by a multidimensional energy landscape, which is determined by the spatial positions of the atoms. Offering unprecedented insights into complex physical and chemical phenomena, molecular dynamics simulation is indispensable in the computational chemistry toolbox. One of the widely recognized issues in MD simulations is the lack of accuracy of underlying interatomic potentials, which hinders truly predictive modeling of molecular systems' properties.
Although the potential energy values could, in principle, be obtained on-the-fly using pertinent electronic structure calculations, more efficient approaches are often required to characterize relevant observables, for example, (ro-)vibrational spectra, rate constants, etc. Thanks to modern computational resources, a large number of highly precise electronic structure calculations can be carried out at various points across the configuration space \cite{Bartlett2007}. Despite significant advances, representing a global PES from these \textit{ab initio} points poses a difficult challenge. Thus, one of the key focuses of computational chemistry has been on the development of efficient methods to generate high-quality representations of global PESs for molecules. Various non-ML approaches, such as modified Sheppard interpolation \cite{Collins2002}, interpolating moving least-squares (IMLS) \cite{Guo2004}, reproducing kernel Hilbert space interpolations \cite{Ho2003} and PIP regression \cite{Qu2018,Houston2022}, to name a few, have been successful at constructing global PESs. Machine learning, particularly neural networks, has recently shown great promise as tools for building both flexible and computationally efficient models of PESs \cite{Behler2007, Schutt2017, Unke2019}. Combining techniques coming from ML and non-ML perspectives has been demonstrated to be effective; PIP-IMLS \cite{Majumder2015} and PIP-NN \cite{Jiang2013, Li2020} are excellent examples of this synergy. 

In the current study, we undertake the development of ML PES for molecular simulations of atmospheric gases with flexible monomers. Collision processes in atmospheric gases modify the radiative properties of moieties by shifting and broadening spectral lines as well as leading to the emergence of collision-induced bands. The advancement of the observational and retrieval capabilities of terrestrial and planetary remote-sensing missions poses new challenges for theoretical research to improve the representation of these phenomena. Starting from potential energy and dipole moment surfaces, contemporary state-of-the-art theories permit first-principle simulations of spectral features that emerge in real gases (see review by \citet{Hartmann2018} and references therein). One of the hurdles affecting remote-sensing retrievals is the lack of or insufficient accuracy of models describing continuum absorption. 
In Refs. \cite{Chistikov2019,Chistikov2021}, we developed a first-principle trajectory-based approach to simulate collision-induced absorption. In light of the approach's effectiveness in simulating far-infrared absorption in the cases of N$_2-$N$_2$ \cite{Chistikov2019}, CO$_2-$Ar \cite{Chistikov2021} and N$_2-$CH$_4$ \cite{Finenko2022}, we seek to simulate the forbidden rotovibrational fundamental and overtone bands of molecule pairs consisting of dipoleless moieties.
Following many-body expansion \cite{Stone1997}, we specifically focus on the full-dimensional representation of the two-body term, $V_{2b}$, which corresponds to the interaction energy between two molecules. The two-body component is defined as the difference between the total energy of the dimer $V_\textrm{dimer}$ and the one-body energies of isolated monomers $V_{\textrm{mon}_1}$ and $V_{\textrm{mon}_2}$
\begin{gather}
    V_\textrm{2b} = V_\textrm{dimer} - V_{\textrm{mon}_1} - V_{\textrm{mon}_2}.
    \label{intro:two-body}
\end{gather}
Note that in this context dimer refers to a molecular pair that is not necessarily in a bound state. Keep in mind that the two-body component of the PES can be paired with various monomer potentials depending on the level of fidelity required for a particular application, making the global PES model configurable and adaptable.

The machine-learning potential is typically built from two components: the molecular descriptor and the ML model. The descriptor provides a mapping that associates an atomic configuration, which is identified by the positions and chemical identities of atoms, with a point in a feature space \cite{Behler2016}. Although the Cartesian coordinates of the atoms encode all the information about the structure of a molecular system, it is evident that they cannot be used directly as input to the ML model. In principle, the potential energy surface is invariant to overall translation and rotation as well as to permutations of identical atoms. Generally speaking, ML models that account for symmetry either using descriptors or through clever architecture tend to be more data-efficient and accurate. In this paper, we use PIPs as a type of descriptor that will allow us to account for translational, rotational, and permutational symmetry.


In recent years, numerous ML methods have been applied to fit electronic energies (see review \cite{Manzhos2021} and references therein). These ML methods produce mappings that can be characterized as parametric or nonparametric \cite{Russell2015}. The assumption made by parametric algorithms is that the mapping function has a predefined functional form with a fixed number of parameters. The typical example is a neural network. Nonparametric algorithms, on the other hand, do not make this assumption. The more training data is given, the more complex the algorithm becomes and the more parameters it contains. Such is the case with kernel methods like Gaussian process regression and kernel ridge regression. Assuming the NN's structure is fixed, the prediction cost for NN-based methods becomes independent of the size of the training set, in contrast to kernel-based approaches where the prediction cost scales linearly with the size of the training set. Given our goal of performing large-scale collision simulations with ML potentials, which necessitates robust force models, we chose an NN-based method.

To assess the computational efficiency of the PIP-NN method, we conducted a showcase study on the ethanol molecule using the MD17 database of energies and forces. A range of ML approaches have been compared by \citet{Pinheiro2021} based on the MD17-ethanol database. Using the suggested protocol based on the analysis of learning curves, we compare our implementation of the PIP-NN method to other approaches, in particular, symmetrized gradient-domain machine learning (sGDML) \cite{Chmiela2018, Sauceda2019}, deep neural network with PhysNet architecture \cite{Unke2019}, Gaussian approximation potential \cite{Bartok2015}, and PIP regression \cite{Houston2022} on the basis of precision and prediction times.


The rest of the paper is organized as follows. The structure of the PIP-NN model and a few modifications that we explore in this paper are described in the next section, along with the generation of interaction energy data sets and an explanation of the training procedure. The Results section discusses how PIP-NN models performed on these data sets as well as on the MD17-ethanol data set. The Summary and Conclusions section covers concluding remarks. 

\section{Methods}
\label{sec:methods}

In this work, we adapt the PIP-NN approach put forward by Guo and co-workers \cite{Jiang2013, Li2013} to represent multidimensional PESs, characterized by two-body, non-covalent interactions. We begin by outlining the PIP-NN model's structure and how the PIP basis could be altered to be able to reproduce essential properties pertinent to the two-body component of the PES. Next, we explain how the variables of invariant polynomials could be changed to more accurately map the long-range region. Finally, we discuss the training procedure and the development of interaction energy data sets.

\subsection{Neural network}
\label{subsec:neural-network}

Dense layers are the basic building blocks of a fully-connected NN. They represent only linear transformations from input to output. It was established that arbitrary multivariate relationships could be modeled via combining the univariate non-linear functions and dense layers \cite{Kolmogorov1957}. A single-hidden layer NN is a mapping
\begin{gather}
    y_k = \sum_{n = 0}^{n_\textrm{hidden}} w^{(2)}_{n, k} \sigma \lb \sum_{j = 1}^{d_\textrm{in}} w^{(1)}_{j, n} x_{j} + b^{(1)}_n \rb + b^{(2)}_k, \quad k = 1 \dots d_\textrm{out},
    \label{methods:single-layer-nn}
\end{gather}
where $\sigma$ is a neuron activation function or simply neuron, $d_\textrm{in}$ and $d_\textrm{out}$ denote the dimensionality of input and output, and $n_\textrm{hidden}$ denotes the number of neurons in the hidden layer.  
The argument of $\sigma$ is a linear form of the coordinates $\mf{x}$ with tunable weights $\boldsymbol{\omega}$ and biases $b$, which are fitted to reproduce a set of training data. 
Building a general multi-layer NN involves feeding the outputs of one layer of neurons into another. 
The architecture of a multi-layer NN could be represented as ($d_\text{in}-\ldots-d_\text{out}$), where the number of neurons in the hidden layers is substituted for the ellipsis.
With the exception of the output layer, where a linear activation function is commonly used, the same activation function is typically used from layer to layer and within the same layer.
A single-layer NN of Eq.~\eqref{methods:single-layer-nn} may approximate any mapping between inputs and outputs assuming it has sufficiently many neurons and an appropriate activation function is used \cite{Hornik1991}, making it a universal approximator \cite{Gorban1998}. There are numerous options for $\sigma$ in use, for example, $\texttt{tanh}(x)$ or $\texttt{max}(0, x)$ \cite{Zeiler2013}. In this work, a Swish activation function \cite{Ramachandran2017} given by $\sigma(x) = x \cdot \texttt{sigmoid}(x)$ is employed.

Since fully-connected NNs recognize no symmetry, additional steps must be taken to incorporate necessary symmetry properties into the model. For molecular systems, it is of particular importance to make sure that the PES model is an invariant of the complete nuclear permutation and inversion group. The enforcement of permutation symmetry is not only crucial for producing correct PESs but also makes them more data-efficient and robust. To this end, Bowman and co-workers have proposed expressing a PES in terms of PIPs of internuclear distances, which guarantees permutational invariance of the model \cite{Braams2009}. Later on, Guo and co-workers used PIPs in the input layer of the fully-connected NN to produce ML PESs of reactive systems \cite{Jiang2013, Li2013}. 

For a system with $N$ atoms, PIPs can be obtained by symmetrizing transformed internuclear distances $\boldsymbol{\tau}$ 
\begin{gather}
    \mathcal{p}_k^\textrm{F} = \mathscr{S} \lsq \prod_{i < j}^{N} \tau_{ij}^{b_{ij}} \rsq, \quad m = \sum_{i < j}^{N} b_{ij},
    \label{methods:pips}
\end{gather}
where $b_{ij}$ is the degree of monomial $\tau_{ij}$ and $m$ is the total degree of PIP. A symmetrization operator $\mathscr{S}$ contains all feasible nuclear permutation operations in the system. Bowman and co-workers developed two approaches to constructing the basis of invariant polynomials of Eq.~\eqref{methods:pips}. In the more sophisticated approach, the polynomials are represented as products of invariant primary and secondary polynomials \cite{Braams2009} that are produced by using a computer algebra system. The second approach is a straightforward monomial symmetrization in which monomials are made invariant by applying all permutations of identical atoms to generate multiterm polynomials \cite{Xie2010}. Because of the simplicity of the approach and the resulting basis set being easy to modify, we chose to employ the monomial symmetrization approach. However, it is important to note that the factorization method derived from invariant polynomial theory must be used instead when the order of the permutation group is large. In contrast to the ``purified'' polynomial bases we describe below, we will refer to the basis of Eq.~(1) as full (F).  

One of the fundamental properties of the two-body potential $V_{2b}$ is that it approaches zero as monomers are separated at large distances. Since not all polynomials in Eq.~\eqref{methods:pips} vanish at large intermonomer separations, the zero-interaction limit can only be reproduced numerically to a limited degree of accuracy. Truhlar and co-workers \cite{Paukku2013} suggested removing those unconnected terms from the invariant basis. The resulting basis is usually referred to as purified (P). In practice, this is achieved by choosing a variety of cluster configurations, setting to zero transformed distances $\boldsymbol{\tau}$ corresponding to intermonomer degrees of freedom, and setting the remaining variables to non-zero values. A purified basis is produced after evaluating each polynomial in the full basis and discarding any that produce non-zero values.

The asymptotic zero-interaction limit of the polynomial basis of Eq.~\eqref{methods:pips} can be alternatively ensured by restricting the basis set to polynomials that are constructed exclusively from the intermonomer transformed distances. This basis will be termed ``pruned purified'' (PP) \cite{Conte2015}. The PP basis set is usually significantly more compact, and its use results in a substantial speed-up in the model evaluation. To determine the elements of the PP basis, we set the intramonomer distance variables to zero while setting the remaining variables to non-zero values. The polynomials producing a value other than zero are maintained to form the PP basis. As we demonstrate below, when a PES employing the rigid-the rotor approximation is required for some applications, the PP basis set is a practical tool for constructing a high-fidelity and robust model.

The purification of the basis set is not only important to achieve the zero-interaction limit but also to improve the precision of the model in the long range. In this asymptotic region, the ML PES should not only uncouple but also do so in a way consistent with the long-range perturbation theory of intermolecular interaction \cite{Avoird1980}. The majority of PIP models utilize Morse variables for transformed internuclear distances, which allows them to qualitatively correctly decay to zero. However, they generally do so faster than true interaction due to exponential-type dependence \cite{Homayoon2015}. This results in large errors in the long range, which can be relevant in certain applications such as low-energy scattering or simulating collision-induced absorption. A remedy suggested in the literature \cite{Homayoon2015, Qu2018} is to construct a two-component fit as a combination of short-range and long-range models. By employing a sufficiently large parameter inside the Morse variable, which permits slower decay, the long-range model closely reproduces correct asymptotic behavior. Although this approach greatly improves the accuracy of the model in the long range region, it also makes the model more complicated and less robust.
Here, we modify the functional form of the transformed distances $\boldsymbol{\tau}$ to improve the long-range behavior of the model. With the help of long-range perturbation theory, one can determine the $R$-dependence of leading terms in the multipole expansion of interaction energy. For this reason, for intermonomer transformed distances, we use a mixed Morse-polynomial variable expressed as
\begin{gather}
    \tau_{ij} = (1 - s) \cdot e^{-r_{ij} / \lambda} + s \cdot \frac{a}{r_{ij}^{n_\text{LR}}}, 
    \label{methods:xi-variable-1}
\end{gather}
where $\lambda$ and $a$ are range parameters, $r_{ij}$ is the internuclear distance between atoms $i$ and $j$, $n_\text{LR}$ is the long-range exponent, and $s$ is a switching function. The range parameter, $\lambda$, is usually chosen to be 2 bohr, and $a$ is set at the appropriate value so that both terms in a mixed variable, Eq.~\eqref{methods:xi-variable-1}, are of the same order in the long range region. In turn, for intramonomer transformed distances, we use a Morse variable $\tau_{ij} = e^{-r_{ij} / \lambda}$. Switching function $s$ smoothly interpolates between the polynomial dependence at long range and Morse-type dependence at short distances. It is defined as
\begin{gather}
    s = \left\{
    \begin{aligned}
        & 0                             && \textrm{for } R < R_\textrm{i}, \\
        & 10 t^3 - 15 t^4 + 6 t^5 \quad && \textrm{for } R_\textrm{i} < R < R_\textrm{f}, \\
        & 1                             && \textrm{for } R > R_\textrm{f},
    \end{aligned}
    \right.
    \label{methods:switching-function}
\end{gather}
where $t = (R - R_\textrm{i}) / (R - R_\textrm{f})$, $R_\textrm{i}$ and $R_\textrm{f}$ are the boundaries of the switching region and $R$ is the distance between the center of mass of two monomers. 

The MSA software \cite{Xie2010} is used to construct the polynomials of the full basis. The procedure implemented in the MSA software involves applying permutations of the identical atoms to a seed monomial. The transformed distances map into another set of variables and thus another monomial, but of the same order. Summing up the monomials produces a permutationally invariant polynomial.
The software accepts the permutational symmetry group and the overall order of the PIPs. Higher-order polynomials are factored by the original program in terms of the lower-order polynomials plus the remainder. Our tests demonstrated that, at least for overall orders $M = 3, 4$, the recursive computation of polynomials is not significantly faster than the straightforward computation for permutational groups of systems in this work. For this reason, we chose a non-recursive approach for computing purified and pruned purified basis sets primarily to make simpler the respective gradient computations.

In the PIP-NN approach, the values of a selected invariant basis set are used as the input vector of NN. The prediction for energy is expressed as a single output neuron with a linear activation function. 
A neural network's parameters must be initialized before training can begin. All the entries of the neural network are initialized with random values uniformly distributed according to the \citet{He2015}'s scheme. The inputs and targets of the NN are scaled to the same range by subtracting the mean values over the training set and divided by the root-mean-square over the training set. The reasons for this are two-fold: first, to roughly align the data range with the parameter initialization range; and second, to make the optimizer's job easier by rescuing it from having to deal with drastically different length scales in various dimensions. In the end, scaling helps in locating a reasonable local minimum of the loss function. After initialization and data normalization, a gradient-based Broyden-Fletcher-Goldfarb-Shanno (BFGS) quasi-Newton optimization procedure \cite{Nocedal1999} tunes the parameters of the neural network to minimize a loss function. We note that the BFGS optimization procedure requires the computation of second derivatives of the NN response, in contrast to the standard backpropagation training algorithm \cite{Rumelhart1986} for NNs, which relies on the computation of only the first derivatives of the NN response. For data sets containing only energies, the loss function is
\begin{gather}
    \mathcal{L} = \frac{1}{N} \sum_{i = 1}^{N} w^{(E)}_i \lb \hat{E}_i - E_i \rb^2.
    \label{methods:loss-energy}
\end{gather}
When quantum-chemically calculated forces are available, one can incorporate them in the training set to increase the accuracy of the model. It has been shown that the PES curvature provided by the forces can substantially improve data efficiency, enabling the model to reach the same target accuracy using fewer data points when training on energies and forces as opposed to training on energies only \cite{Pinheiro2021}. However, the additional cost of performing force calculations quantum-chemically must be taken into account. State-of-the-art methods, such as the coupled-cluster approach, could take two to three times as long to compute forces as they do to compute energy, while some standard DFT implementations can do so with little additional expense. We modify the loss function to include gradient information as follows:
\begin{gather}
    \mathcal{L} = \frac{1}{N_\textrm{c}} \sum_{i = 1}^{N_\textrm{c}} w^{(E)}_i \lb \hat{E}_i - E_i \rb^2 + \frac{1}{N_\textrm{c}} \sum_{i = 1}^{N_\textrm{c}} \frac{w^{(F)}_i}{3 N_\textrm{at}} \sum_{\alpha = 1}^{N_\textrm{at}} \lb \hat{\mf{F}}_{i, \alpha} - \mf{F}_{i, \alpha} \rb \lb \hat{\mf{F}}_{i, \alpha} - \mf{F}_{i, \alpha} \rb^\top.
    \label{methods:loss-forces}
\end{gather}
Here, $N_\textrm{c}$ and $N_\textrm{at}$ are the number of configurations and atoms, $E_i$ and $\mf{F}_{i, \alpha}$ are the reference energies and Cartesian components of the reference force acting on atom $\alpha$ for the $i$-th configuration, respectively, and $\hat{E}_i$ and $\hat{\mf{F}}_{i, \alpha}$ are the predicted energy and atomic force values. The latter are calculated analytically by determining the predicted energy gradient with respect to the atoms' Cartesian coordinates (see details below). 
The weighting hyperparameters $w(E)$ and $w(F)$ determine how much each error term contributes to the loss. Since \textit{ab initio} calculations encompass a wide range of energies, it is beneficial to unequally weight the points in the loss function to improve the accuracy in more valuable energy regions by simultaneously pushing the errors to less valuable regions.
We found that the following weight scheme performs well as a default for both $w^{(E)}$ and $w^{(F)}$
\begin{gather}
    w = \frac{\Delta}{\Delta + E_i - E_\text{min}}.
    \label{methods:weights}   
\end{gather}
Here, $E_\text{min}$ denotes the lowest energy in the training set and hyperparameter $\Delta$ determines the range of favorably weighted energies, which was set to 1000~cm$^{-1}$ for the cases reported in this study after careful consideration.

We randomly divide the total data set into a training, a validation, and a test set, and we regard the error on the test set as an indicator of the overall accuracy of the ML PES. This is done because NNs are prone to overfitting, which means that they achieve a better match of the fitted data at the expense of the poorer quality in between the fitting points. An error on the validation set is computed at each step of the optimization routine. However, instead of using the error to guide parameter optimization, fitting is halted when the error starts increasing (so-called early stopping \cite{Montavon2012}). During training, the model state is continuously saved. This enables us to select the model with the best overall validation loss after training is finished. We use conventional learning rate decay to enhance convergence near the optimal error point. We found that, typically, the model reaches a competitive state within several hundred epochs and most of the remaining training time is spent on only small improvements in the errors. In addition, we observed that increasing the number of layers had little to no impact on the model's overall accuracy, indicating that it is not an important hyperparameter in our tests.

The PIP-NN models are constructed and trained using the PyTorch deep learning framework \cite{PyTorch} on a NVIDIA QuADro GV100 in single-GPU mode and \texttt{float32} precision. After the fitting is complete, the model state is exported in the zipped format and can be loaded for inference in the C++11 code that implements the PIP-NN architecture around the Eigen linear algebra library \cite{Eigen}.

\subsection{Gradient evaluation}
\label{subsec:gradient}

Let us discuss the algorithmic approach to calculating the analytical gradients of the PIP-NN model. The following steps make up a computational plan for predicting energy. Starting with the Cartesian coordinates of atoms, $\mf{x}$, we compute the corresponding internuclear distances, $\mf{r}$, transform them into $\boldsymbol{\tau}$ of either Morse or mixed Morse-polynomial type, and then use the latter as variables for invariant polynomials, $\boldsymbol{\mathcal{p}}$, whose values are eventually passed to NN to produce the prediction $\hat{E}$. We invoke a chain rule on the described computational plan to investigate individual terms in line with the algorithmic differentiation approach
\begin{gather}
    \frac{\partial \hat{E}}{\partial \mf{x}} = \frac{\partial \hat{E}}{\partial \boldsymbol{\mathcal{p}}} \frac{\partial \boldsymbol{\mathcal{p}}}{\partial \boldsymbol{\tau}} \frac{\partial \boldsymbol{\tau}}{\partial \mf{r}} \frac{\partial \mf{r}}{\partial \mf{x}}.
    \label{methods:energy-gradient}
\end{gather}

To compute the first term of Eq.~\eqref{methods:energy-gradient}, we begin at the first hidden layer of the neural network and obtain the derivatives of its outputs with respect to its inputs. Then, by expressing the following layer in terms of the preceding one, we recursively differentiate the output with respect to the $i$-th element of the PIP basis set
\begin{gather}
    \begin{aligned}
        \frac{\partial \mf{H}^{(k)}}{\partial \mathcal{p}_i} \Bigg\vert_{q = 1 \ldots n_k} &= \lsq \frac{\partial \sigma \lb \mf{W}^{(k)}_{q} \cdot \mf{H}^{(k - 1)} + b^{(k)}_{q} \rb}{\partial \mathcal{p}_i} \rsq_{q = 1 \ldots n_k} \\  
        &= \lb \mf{W}^{(k)}_{q} \cdot \frac{\partial \mf{H}^{(k - 1)}}{\partial \mathcal{p}_i} \rb \frac{\partial \sigma}{\partial \mathcal{p}_i} \lb \mf{W}^{(k)}_{q} \cdot \mf{H}^{(k - 1)} + b^{(k)}_{q} \rb \Bigg\vert_{q = 1 \ldots n_k},
    \end{aligned}
    \label{methods:recursive-gradient}
\end{gather}
where $\mf{W}^{(k)}$, $b^{(k)}$ and $\mf{H}^{(k)}$ denote the weight matrix, bias and the output vector of the $k$-th hidden layer of size $n_k$ . We acquire the derivative of NN's output with respect to its inputs by recursively inserting the results for subsequent layers back into Eq.~\eqref{methods:recursive-gradient}.

The second term, $\partial \boldsymbol{\mathcal{p}}/\partial \boldsymbol{\tau}$, is essentially a Jacobian matrix for a vector function that associates transformed distances with elements of the PIP basis set. 
Using a combination of text and expression manipulation, we first procure symbolic expressions for Jacobian matrix elements, which are subsequently converted into procedural code. Since we avoid recursive factoring of the PIP basis set and compute each of its elements separately, the symbolic differentiation technique for the Jacobian matrix elements is effective and straightforward. The MSA software was extended with the implementation of this technique in Python. For any basis set type (P, PP, or F) given in the dedicated format suggested in the MSA software, the program generates C code for evaluating the Jacobian matrix in addition to C code for evaluating the PIP basis set's elements. It goes without saying that the program needs to be run for each basis set type and permutational symmetry group in which the user could be interested.
    
The third and fourth terms of Eq.~\eqref{methods:energy-gradient} are computed using expressions produced via symbolic differentiation. After all the terms have been multiplied as matrices, which is notably the most computationally demanding step, the derivatives of energy with respect to Cartesian coordinates are obtained.

\subsection{Data set generation}
\label{subsec:dataset-generation}

In the following, we describe the generation of two data sets that probe intermolecular interactions for N$_2-$Ar and N$_2-$CH$_4$.
For each dimer configuration, the energies of the dimer and the isolated monomers at that configuration were calculated, and the intrinsic two-body energy was computed according to Eq.~\ref{intro:two-body} while accounting for the basis set superposition error using the Boys and Bernardi counterpoise correction scheme \cite{Boys1970}. The atomic forces were not calculated due to their large computational cost within the supermolecular approach.

\subsubsection{N$_2-$Ar}
\label{subsec:n2-ar}

We used Jacobi coordinates ($R$, $l_{\text{N}_2}$, and $\theta$) to construct a grid of configurations for the N$_2-$Ar complex, where $R$ is the distance between moieties' centers of mass, $\theta$ is the enclosed angle between the molecular axis of N$_2$ and the intermolecular vector $\mf{R}$, and $l_{\text{N}_2}$ is the N$_2$ bond length. The grid contains 31 intermolecular distances ranging from 4.5 to 30.0 bohr, 8 points in angle $\theta$ located at the roots of the Legendre polynomial of 8-th order, and 3 intradiatomic distances $l_{\text{N}_2} = [1.97, 2.078, 2.2]$~bohr, yielding a total of 744 configurations.

The single- and double-excitation coupled-cluster method with a noniterative perturbation treatment of triple excitations (CCSD(T)), as implemented in the MOLPRO 2010 program package \citep{MOLPRO}, was adopted to determine the energies. The interaction energies were calculated with a series of correlation-consistent basis sets aug-cc-pVXZ (X = T, Q, 5) \citep{Dunning1989} and extrapolated to the complete basis set limit using a scheme of \citet{Peterson1994}. 

\subsubsection{N$_2-$CH$_4$}
\label{subsec:ch4-n2}

The following steps were taken to create the database of 98\,268 configurations and energies. We employed a collection of 71\,610 points that assume rigidity of both moieties calculated in our previous effort \cite{Finenko2021}. Hereafter, it will be referred to as D1. In this collection, the CH$_4$ and N$_2$  were fixed at zero-point vibrationally-averaged structures characterized by $r(\text{C}-\text{H}) = 2.067$ bohr \cite{Albert2009} and $r(\text{N}-\text{N}) = 2.078$ bohr \cite{Bendtsen2000}, both determined from experimental data. The additional newly calculated interaction energies at 26\,658 configurations for which both monomers are vibrationally excited were compiled into another collection, denoted as D2.
To sample the dimer configurations, a two-step technique based on the density 
\begin{gather}
    \rho \sim \exp \lb -\beta_\text{eff}^{\text{CH}_4} V_{\text{CH}_4} \rb \cdot \exp \lb -\beta_\text{eff}^{\text{N}_2} V_{\text{N}_2} \rb
    \label{methods:sampling-density}
\end{gather}
with effective reciprocal temperatures $\beta_\text{eff}^{\text{CH}_4}$ and $\beta_\text{eff}^{\text{N}_2}$, both set to 500~K, and monomer potentials $V_{\text{CH}_4}$ and $V_{\text{N}_2}$ was employed. First, we independently sampled the Cartesian coordinates for each of the moieties from the density driven by their respective monomer potential; second, we arranged the moieties in the space-fixed system so that CH$_4$'s center of mass coincided with the origin and N$_2$'s center of mass had the following components:
\begin{gather}
    \mf{R} = R 
    \begin{bmatrix}
        \cos \Phi \sin \Theta \\
        \sin \Phi \sin \Theta \\
        \cos \Theta
    \end{bmatrix},
    \label{methods:intermolecular-vector}
\end{gather}
where $\Phi$ and $\cos \Theta$ are distributed uniformly in the range of $[0, 2 \pi]$ and $[0, 1]$, respectively, and intermolecular distance is distributed with density $R^3$. The intermolecular distance $R$ was selected in the range from 4.5 to 30 bohr.  
The Markov chain Monte Carlo (MCMC) sampling procedure was designed to generate probable sequences of the moieties' configurations represented in terms of Cartesian coordinates. Specifically, we take advantage of an ensemble MCMC sampler proposed by \citet{Goodman2010}, which is known to produce independent samples with lower correlation lag compared to conventional MCMC samplers for a variety of distributions.
For CH$_4$, we used a highly accurate nine-dimensional PES due to Tennyson and co-workers \cite{Owens2016}, whereas the monomer potential for N$_2$ was represented as a Morse curve
\begin{gather}
    V_{\text{N}_2}(l_{\text{N}_2}) = D_\text{e} \lb 1 - \exp \lb -a (l_{\text{N}_2} - l_{\text{N}_2}^\text{e}) \rb \rb, \quad a = 2 \pi c \omega_\text{e} \sqrt{\frac{\mu}{2 D_\text{e}}},
    \label{methods:n2-morse}
\end{gather}
where $D_\text{e} = 9.91$~eV is the dissociation limit, $\omega_\text{e} = 2358.57$~cm$^{-1}$ is the harmonic vibrational frequency, and $l_{\text{2}}^\text{e} = 1.09768$~\AA\, is the equilibrium bond length. 

Interaction energies up to several eV were taken into account. This guarantees that model encompasses regions that may be explored by the collision simulations at near room temperature. The monomer energies were bound by 1\,000~cm$^{-1}$ and 3\,000~cm$^{-1}$ for N$_2$ and CH$_4$, respectively. The electronic structure calculations were performed using the explicitly-correlated CCSD(T)-F12a method \cite{Adler2007, Knizia2009} with the aug-cc-pVTZ basis set in the MOLPRO 2010 program package. The geminal exponent $\beta$ in the correlation factor F12 was set to 1.3.  The density functions CABS(OptRI) and standard auxiliary basis sets were employed.
By subtracting the asymptotic energy estimated at $R = 1000$~bohr, the interaction energy for each configuration was adjusted to account for the size inconsistency caused by the use of scaled triple contribution to F12 energy.

\section{Results}
\label{sec:results}

First, we describe the construction and performance of PIP-NN two-body PESs for N$_2-$Ar and N$_2-$CH$_4$, characterized by non-covalent interactions. Using the N$_2-$CH$_4$ system as an example, we demonstrate that the PIP-NN model, which is based on the pruned purified set of polynomials, can perform effectively within the rigid-rotor approximation, outperforming the regression model developed in Ref.\cite{Finenko2021}. The use of mixed Morse-polynomial variables in the flexible monomer scenario is shown to improve the accuracy of the ML PES in the long-range region.
The temperature dependences of the second virial coefficient for N$_2-$Ar and N$_2-$CH$_4$ were then derived, fully accounting for molecular flexibility effects. The fidelity of the PIP-NN models was attested to by a reasonable agreement between the calculated and experimental values. Finally, we undertake a performance analysis of the PIP-NN approach utilizing the protocol put forward by \citet{Pinheiro2021} on the MD17-ethanol database.

\subsection{N$_2-$Ar}
\label{subsec:results-n2-ar}

\begin{figure}[H]
    \centering
    \includegraphics[width=0.7\linewidth]{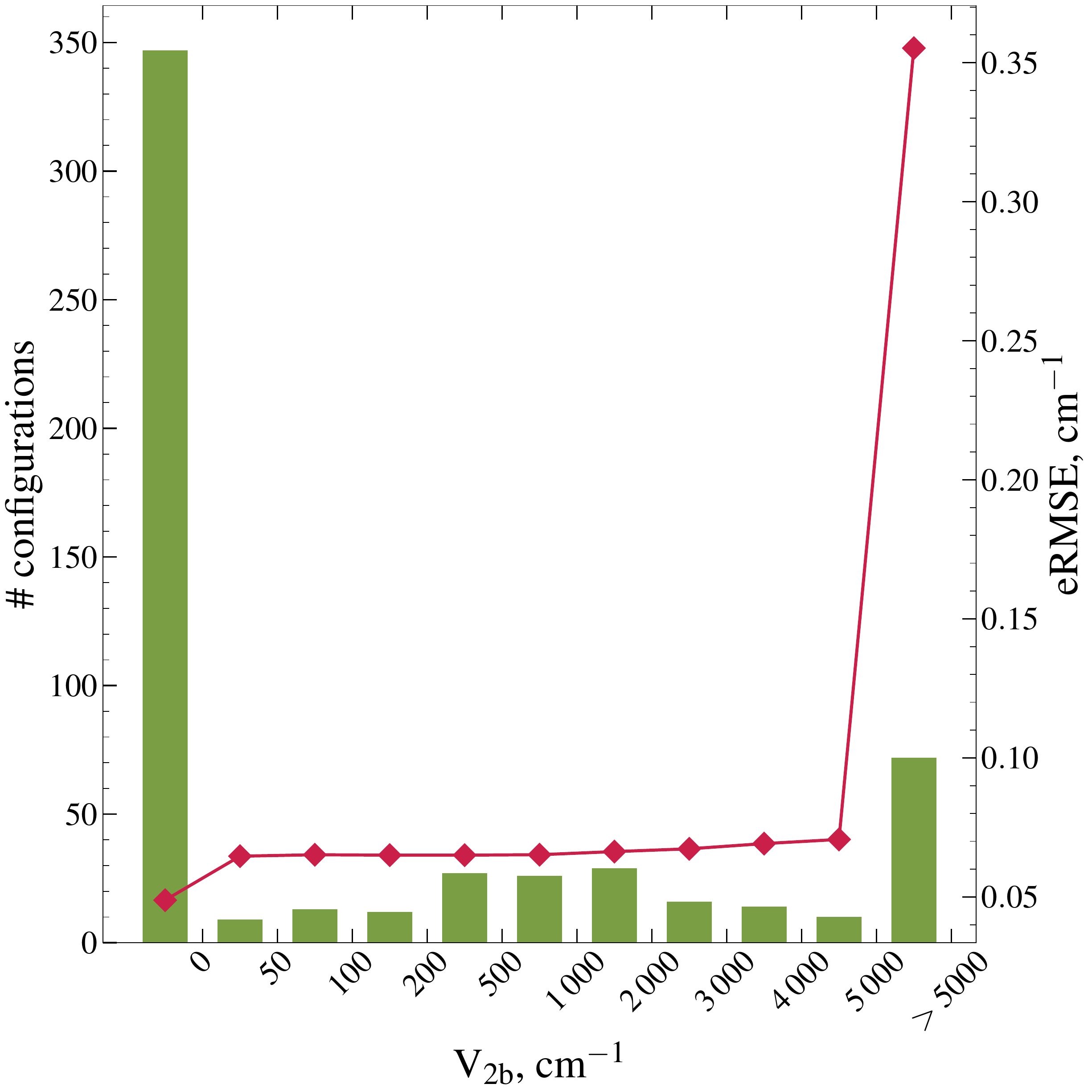}
    \caption{Energy distribution of the 744 points in N$_2-$Ar interaction energy database and cumulative eRMSE of the PIP-NN model.}
    \label{fig:n2-ar-cumulative}
\end{figure}

The purified basis set, with a maximum order of 4, composed of 18 polynomials invariant to 21 symmetry group was first obtained by eliminating unconnected terms from the full basis set. Since dispersion contribution is the leading term in interaction energy for N$_2-$Ar, we set the long-range exponent $n_\text{LR}$ in mixed Morse-polynomial variables for invariant polynomials to 6. The polynomials of the P basis set were used in the input layer of the PIP-NN model of architecture (18-32-1). The database of N$_2-$Ar interaction energies consisting of 744 points was split into training, validation, and testing data sets in the 80–10–10\% proportion. The energy distribution and the cumulative eRMSE error of the PIP-NN model are shown in Figure~\ref{fig:n2-ar-cumulative}. Over the range of interaction energies up to 40\,000~cm$^{-1}$, the model attains an eRMSE of 0.35~cm$^{-1}$.
The PES is characterized by a global T-shaped minimum located at $R = 3.715\,$\AA$\,$, and has a well-depth of $97.3$ cm$^{-1}$. When compared to empirical PESs \cite{Dham1995} and accurate coupled-cluster studies \cite{Munteaunu2004, Candori1983}, the PIP-NN model recovers over 95\% of the well-depth and predicts the minimum location within 0.1\AA.

\subsection{N$_2-$CH$_4$}
\label{subsec:results-n2-ch4}

\begin{figure}
    \centering
    \includegraphics[width=0.7\linewidth]{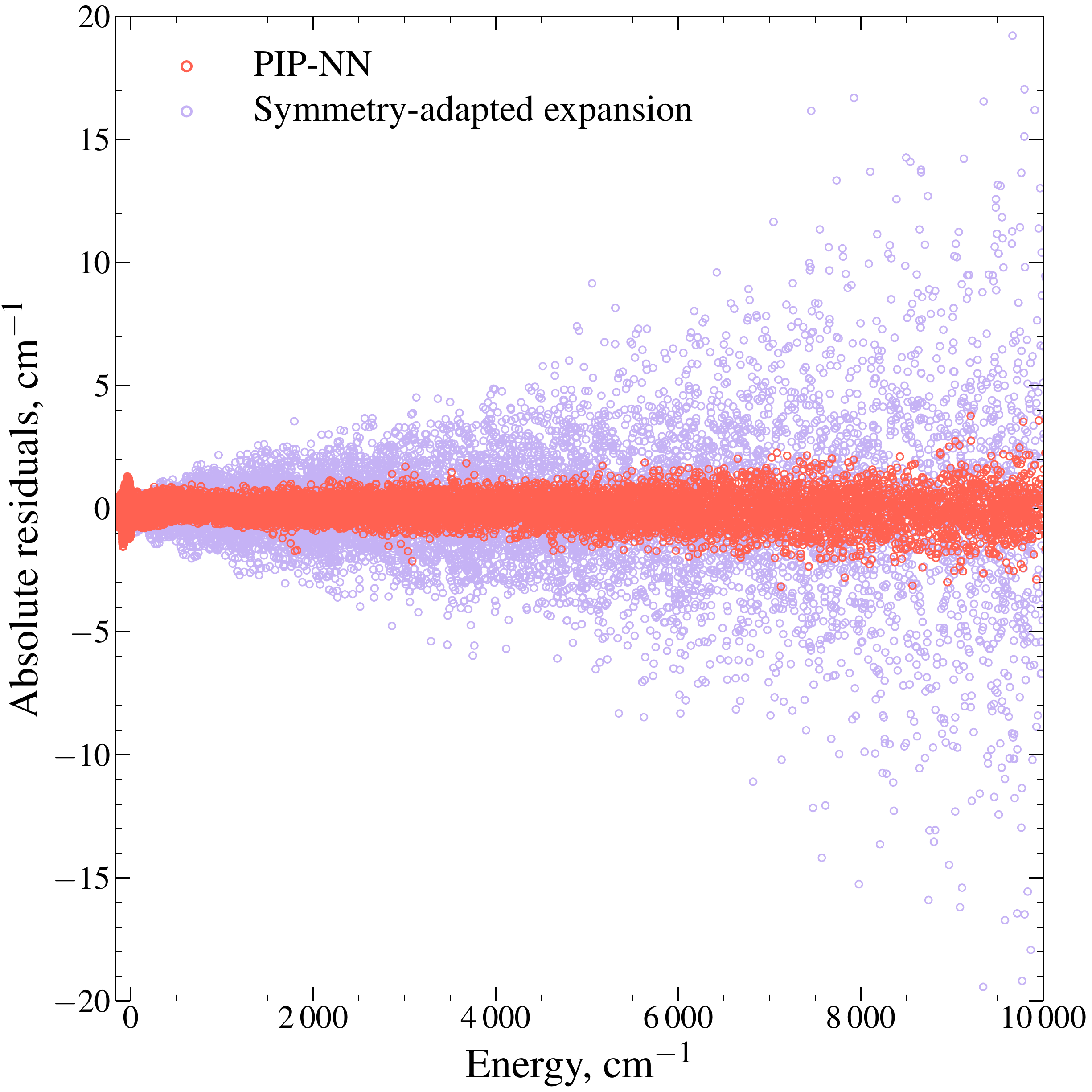}
    \caption{Absolute residuals in the N$_2-$CH$_4$ interaction two-body energy for PES models within rigid-rotor approximation. The PIP-NN model and symmetry-adapted expansion (see text) were constructed based on the D1 data set.}
    \label{fig:n2-ch4-rigid-error-distribution}
\end{figure}

In our previous effort \cite{Finenko2021}, we took advantage of a linear expansion in symmetry-adapted angular functions to construct a high-fidelity N$_2-$CH$_4$ PES within the rigid-rotor approximation. Here, we take this PES as a baseline for the PIP-NN approach. Our objective is to demonstrate that, by utilizing the PIP-NN approach, we can build models with at least the same level of accuracy as commonly used expansions in angular functions derived from the long-range perturbation theory of intermolecular interaction. 
Due to the nature of rigid-rotor approximation, it is reasonable to build the PIP-NN model on top of a PP basis set. To construct the latter, we employ the permutational symmetry group 421 and obtain 650 polynomials of the full basis set with an overall order of 4. The terms that depended on the intramonomer variables were subsequently eliminated, leaving the basis set with only 78 polynomials. We observe the drastic reduction in the basis size brought about by pruning.
The long-range exponent $n_\text{LR}$ was set to 6 because the electrostatic component, which appears as a leading term in the long-range expansion, decays as R$^{-6}$ \cite{Finenko2021}. Using the D1 data set, the PIP-NN model of the (78-32-1) architecture was fitted.
Figure~\ref{fig:n2-ch4-rigid-error-distribution} shows that the PIP-NN model achieves lower residual errors in a wide energy range than the baseline symmetry-adapted expansion of Ref. \cite{Finenko2021}.

\begin{figure}
    \centering
    \includegraphics[width=0.7\linewidth]{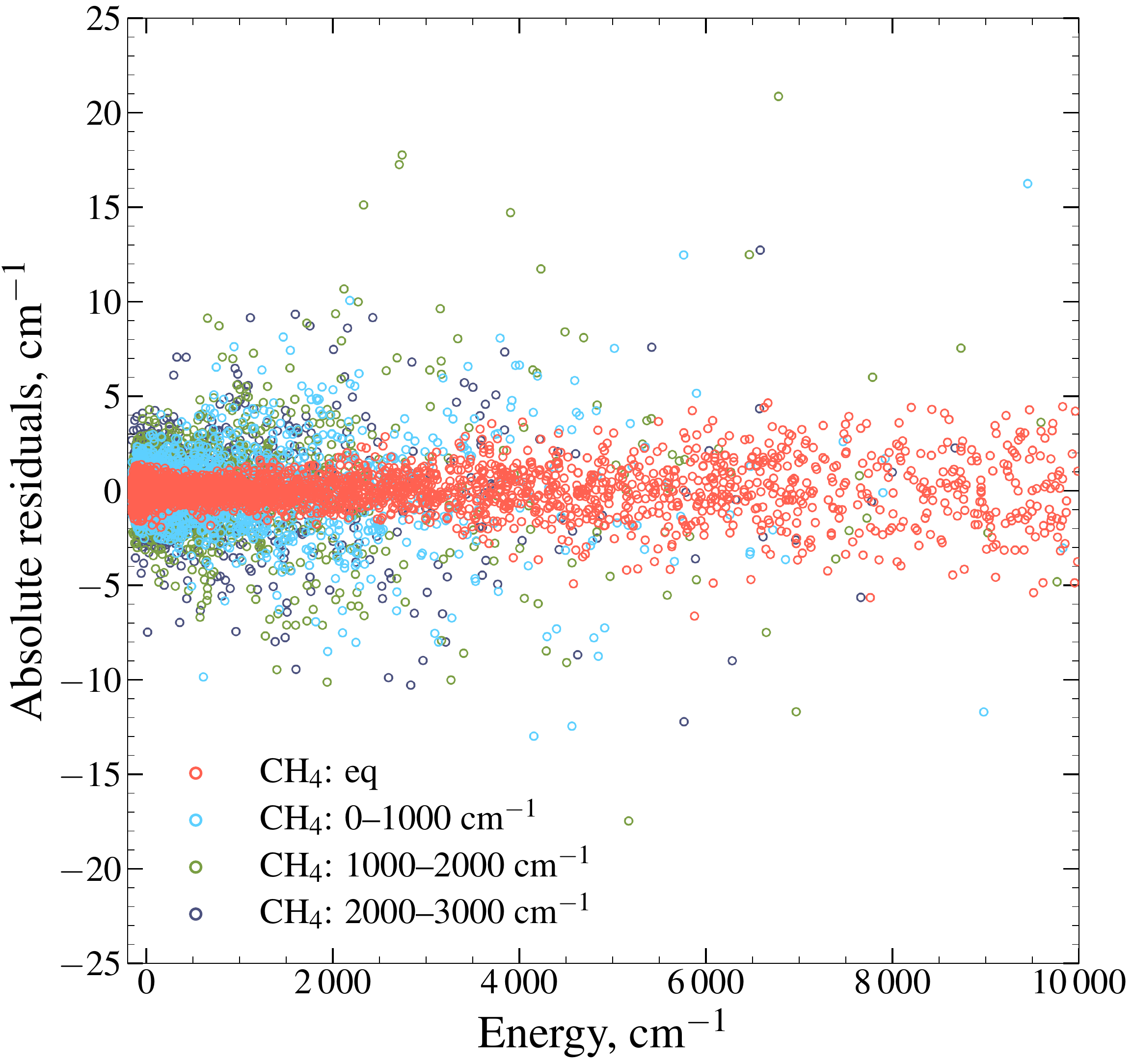}
    \caption{Absolute residuals in N$_2-$CH$_4$ interaction two-body energy obtained for the full-dimensional PIP-NN model. Based on the monomer energy of CH$_4$, the configurations are binned into four ranges: equilibrium (eq), 0$-$1000, 1000$-$2000, and 2000$-$3000~cm$^{-1}$.}
    \label{fig:n2-ch4-nonrigid-overview}
\end{figure}

Given the PIP-NN model's remarkable performance in the rigid-rotor scenario, we proceeded to construct a full-dimensional representation of the two-body interaction energy for N$_2-$CH$_4$. The training data set was obtained by randomly selecting 10,000 configurations from the D1 collection and combining them with the entirety of the D2 collection in order to mitigate the imbalance towards configurations for which both moieties are in their respective equilibrium structures, which could lead to unstable convergence during training. The purified basis must be employed in order to capture the interaction between intermolecular and intramolecular variables. We obtain the 524 polynomials of the purified basis by removing unconnected terms from the full basis set previously obtained. The transformed interatomic distances were the same as those used in the rigid-rotor case. The union of the D1 and D2 data sets mentioned above was used to train the PIP-NN model with the (524, 32, 1) architecture. The absolute residuals in interaction energy for the obtained PIP-NN model are shown in Figure~\ref{fig:n2-ch4-nonrigid-overview}. 
Based on the distribution of residuals and associated eRMSE values, we observe that the accuracy of the prediction of interaction energy declines as the CH$_4$ monomer energy rises. Nevertheless, the model is very accurate with a cumulative eRMSE of 0.85~cm$^{-1}$. 

\begin{figure}
    \centering
    \includegraphics[width=0.7\linewidth]{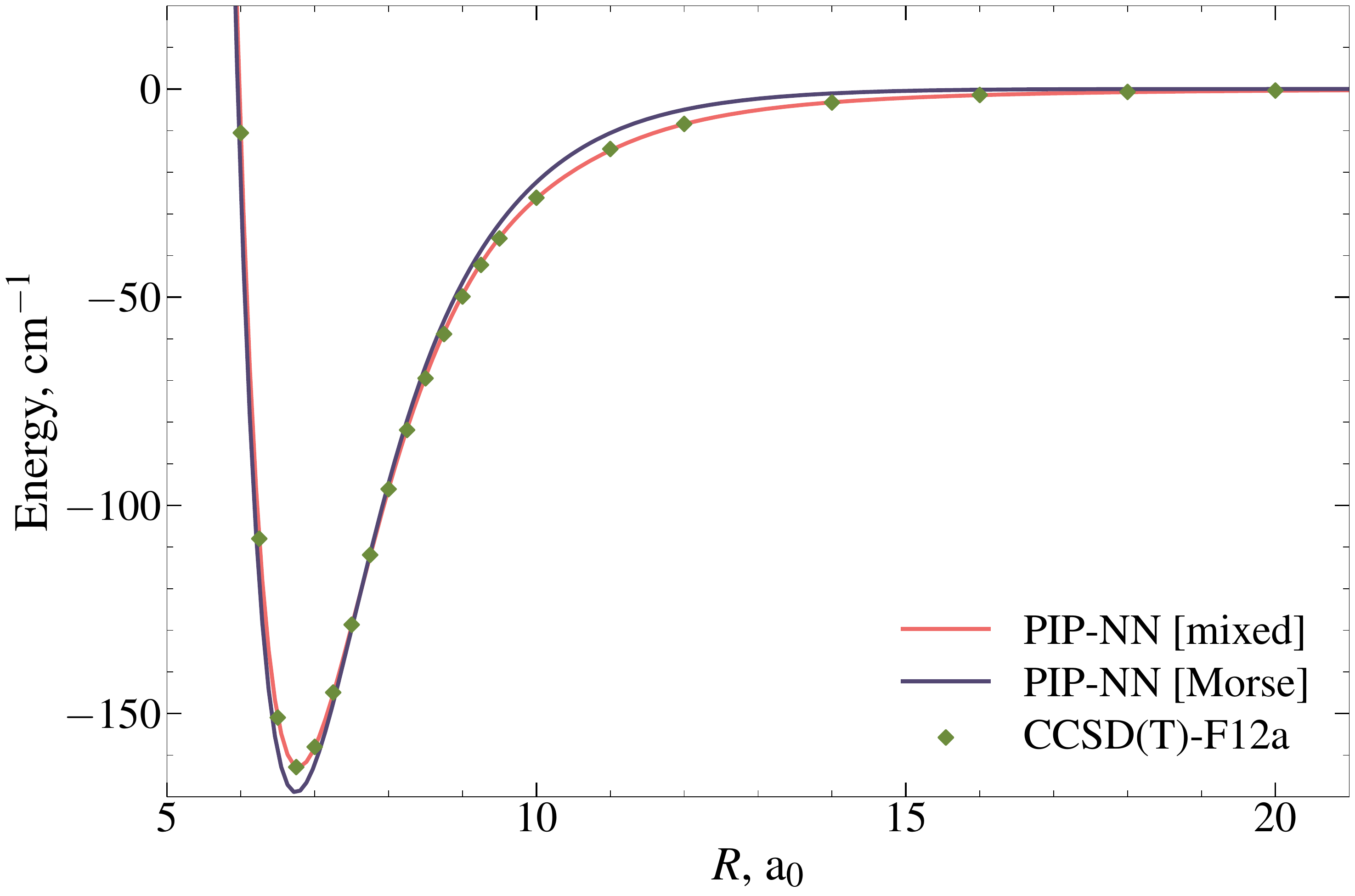}
    \caption{N$_2-$CH$_4$ interaction two-body energy as a function of intermolecular distance. Both moieties taken in their respective equilibrium structures and oriented so that the global minimum over the intermolecular variables is obtained on the one-dimensional cut. The full-dimensional PIP-NN models employ the same purified basis set but have different variable structures, while both having a (524-32-1) architecture. For the intermolecular and intramolecular degrees of freedom in one model (referred to as "PIP-NN mixed"), mixed Morse-polynomial and Morse variables are used, whereas in the other (referred to as "PIP-NN Morse"), Morse variables are used for all degrees of freedom.}
    \label{fig:n2-ch4-potential-cut}
\end{figure}

We further examined the PIP-NN model's accuracy by predicting energies for potential cuts in intermolecular distance, as shown in Figure~\ref{fig:n2-ch4-potential-cut}. Here, both of the monomers are fixed in their respective equilibrium structures and oriented in such a way to produce a cut that has a global minimum over intermolecular coordinates. We trained two PIP-NN models of the (524-32-1) architecture employing the same purified basis set but with different variable structures. 
The first one uses mixed Morse-polynomial and Morse variables for the intermolecular and intramolecular degrees of freedom, while the second one uses Morse variables for both types of degree of freedom. As intermolecular distance increases, the PIP-NN model that utilizes Morse variables for intermolecular degrees of freedom decays too quickly, as seen in Figure~\ref{fig:n2-ch4-potential-cut}. This issue could be mitigated by the use of mixed Morse-polynomial variables. The PIP-NN model utilizing the latter closely replicates quantum-chemical data at both potential well and long range.

\subsection{Second Virial Coefficients}
\label{subsec:svc}

\begin{figure}
    \centering
    \includegraphics[width=0.7\linewidth]{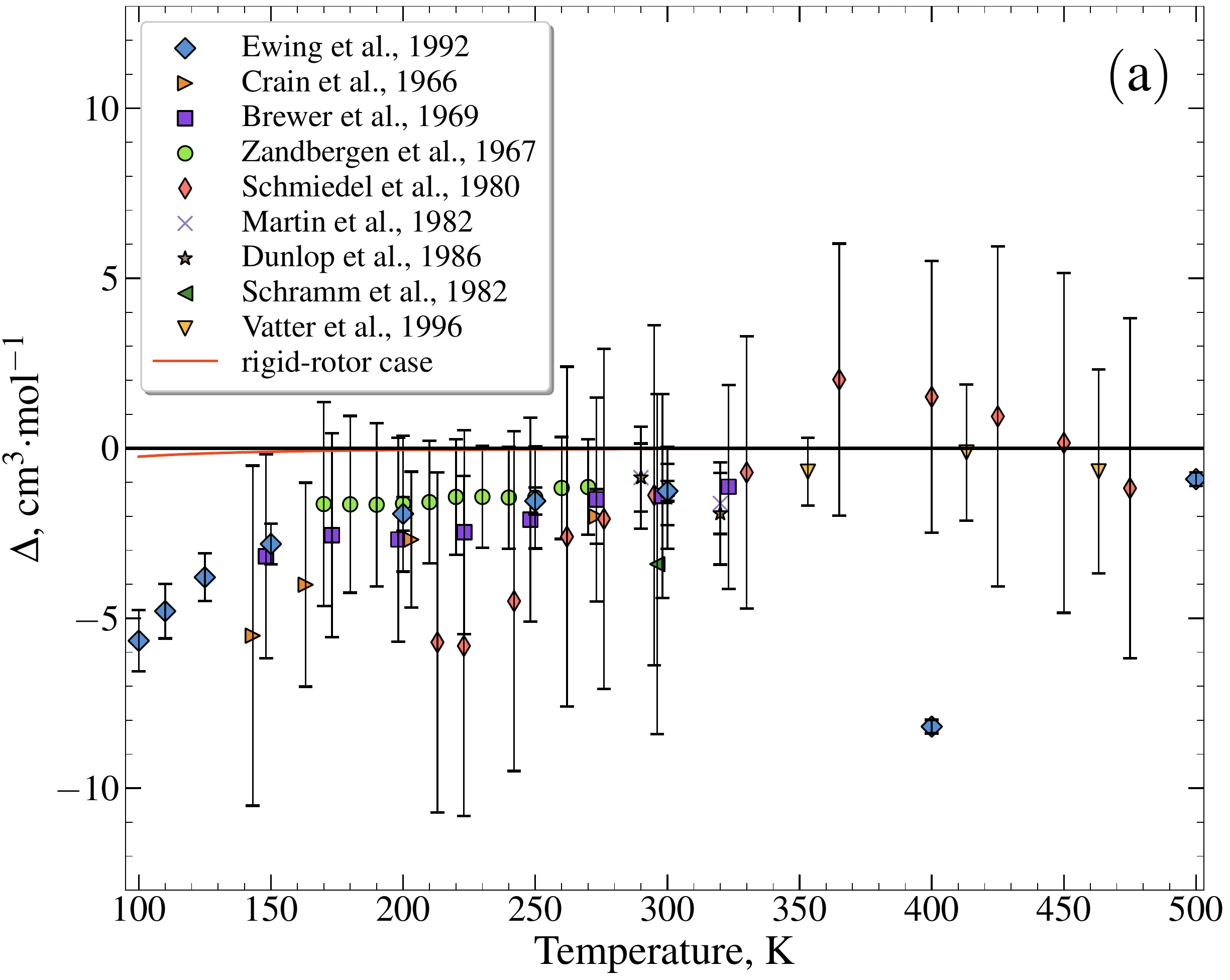}
    \includegraphics[width=0.7\linewidth]{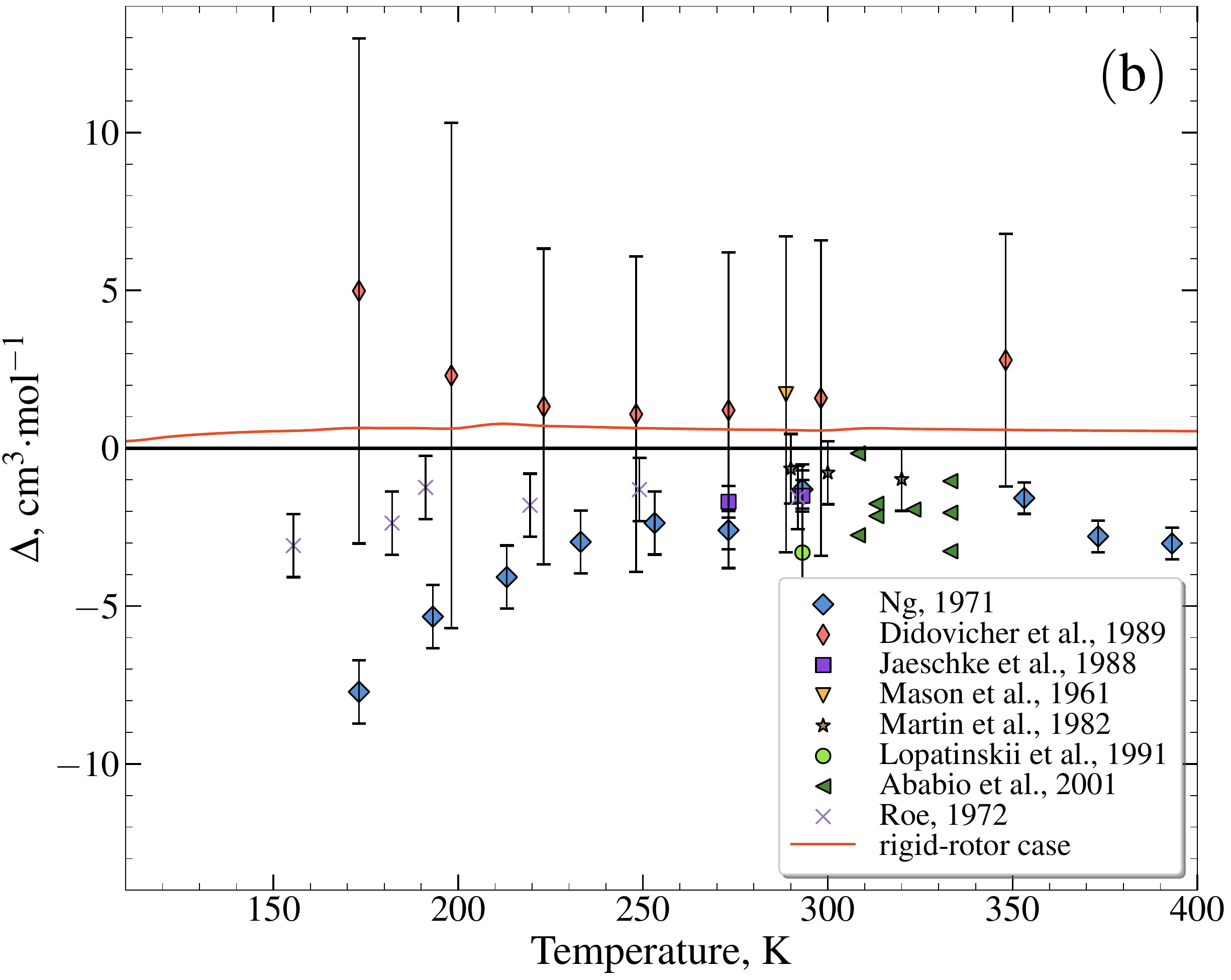}
    \caption{Deviations in N$_2-$Ar (a) and N$_2-$CH$_4$ (b) second virial coefficient, $\Delta = B_{12}^\text{exp} - B_{12}$, of experimental values from the calculated values obtained using Eq.~\eqref{svc:svc}. The orange line represents the difference between the values obtained using the rigid-rotor approximation and those obtained using the full-dimensional PES.}
    \label{fig:svc}
\end{figure}

In classical approximation, the second virial coefficient, $B_{12}(T)$, for flexible monomers is given by \cite{Refson1987}
\begin{gather}
    B_{12}(T) = \frac{V N_\text{A}}{4 \pi} \frac{\displaystyle \int f_{12} \exp \lb -\beta V_1 \rb \exp \lb -\beta V_2 \rb \text{d} \mf{r}^1 \text{d} \mf{r}^2}{\displaystyle \int \exp \lb -\beta V_1 \rb \text{d} \mf{r}^1  \int \exp \lb -\beta V_2 \rb \text{d} \mf{r}^2},
    \label{svc:svc}
\end{gather}
where $f_{12}$ is the Mayer function, $N_A$ is Avogadro number, $V$ is gas volume, and $\mf{r}^1$ and $\mf{r}^2$ denote moieties' Cartesian coordinates. Keep in mind that the integral in the numerator of Eq.~\eqref{svc:svc} presupposes the fixed position of the dimer center of mass. We address the Monte Carlo integration method of Eq.~\eqref{svc:svc} adhering to the importance sampling approach with proposal density prescribed by the denominator. Using the ensemble MCMC approach, thermally distributed monomer configurations were generated and subsequently positioned and oriented with respect to one another as outlined in the Data set generation subsection.
Although the calculation is reasonably inexpensive at a single temperature, performing it repeatedly at various temperatures requires significant processing time. The bottleneck of this computational scheme turns out to be the generation of configurations using the ensemble MCMC method. Provided the points generated at the reference temperature could be reused to obtain an estimate for a temperature at variance with the reference one, significant amounts of processing time could be saved. This could be achieved through the use of the re-weighting factors. Taking into account the temperature variation of proposal density, the re-weighting factors are given by
\begin{gather}
    w_{T_\text{ref}}^{T_\text{target}}(\mf{r}^1, \mf{r}^2) = \frac{\rho \lb \mf{r}^1, \mf{r}^2; T_\text{target} \rb}{\rho \lb \mf{r}^1, \mf{r}^2; T_\text{ref} \rb},
    \label{svc:weighting-factors}
\end{gather}
where $T_\text{ref}$ and $T_\text{target}$ are reference and target temperatures, respectively. The sum of re-weighting factors is an effective number of samples for the target temperature $T_\text{target}$ and should be used for proper averaging. Using the outlined re-weighting technique, we obtained Monte Carlo estimates of second virial coefficients for N$_2-$Ar and N$_2-$CH$_4$ pairs in the temperature ranges 100-500 K and 150-400 K, respectively. Figure~\ref{fig:svc} shows the deviations between the calculated values for N$_2-$Ar and N$_2-$CH$_4$ and the available experimental data (Refs. \cite{Zandbergen1967, Vatter1996, Brewer1969, Dunlop1986, Schmiedel1980, Crain1966, Schramm1982, Ewing1992, Martin1982} for N$_2-$Ar and Refs.\cite{Ng1971, Ababio2001, Roe1972, Martin1982, Jaeschke1988, Didovicher1989, Lopatinskii1991, Mason1961} for N$_2-$CH$_4$). We also show the values of the second virial coefficient that were obtained using the rigid-rotor approximation by conducting integration in Eq.~\ref{svc:svc} over the intermolecular variables only and foregoing the integration over the moieties' intramolecular degrees of freedom.

\subsection{Ethanol}
\label{subsec:results-ethanol}

As we mentioned previously, the MD17 database of energies and forces was used as a playground for multiple ML approaches. The configurations compiled into a data set originate from the \textit{ab initio} MD trajectories calculated at a temperature of 500 K using the van der Waals-corrected density functional theory with the PBE functional (PBE + vdW-TS) \cite{Perdew1996, Tkachenko2009}. 

\begin{figure}
    \centering
    \includegraphics[width=0.49\linewidth]{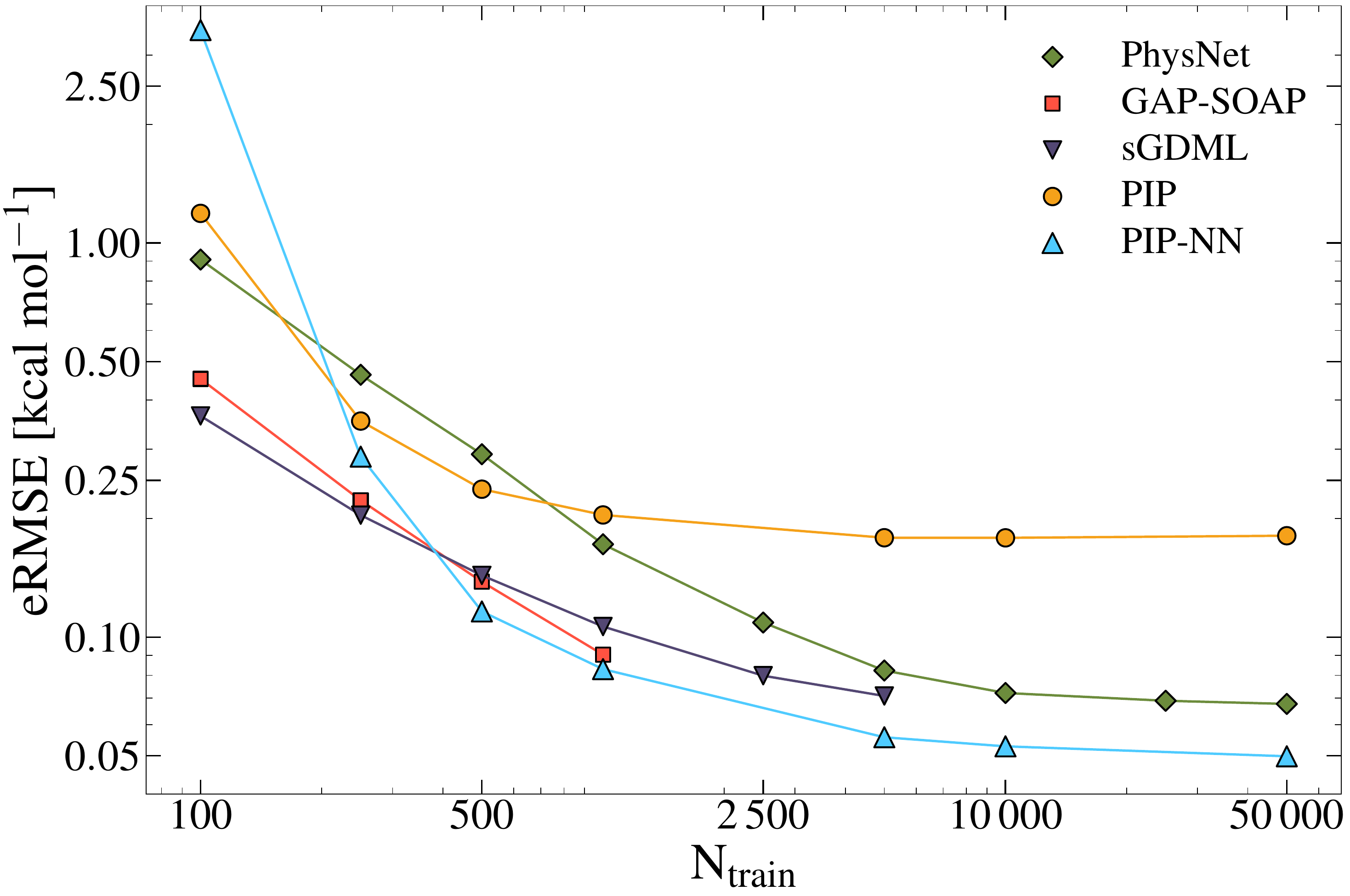}
    \hfill
    \includegraphics[width=0.49\linewidth]{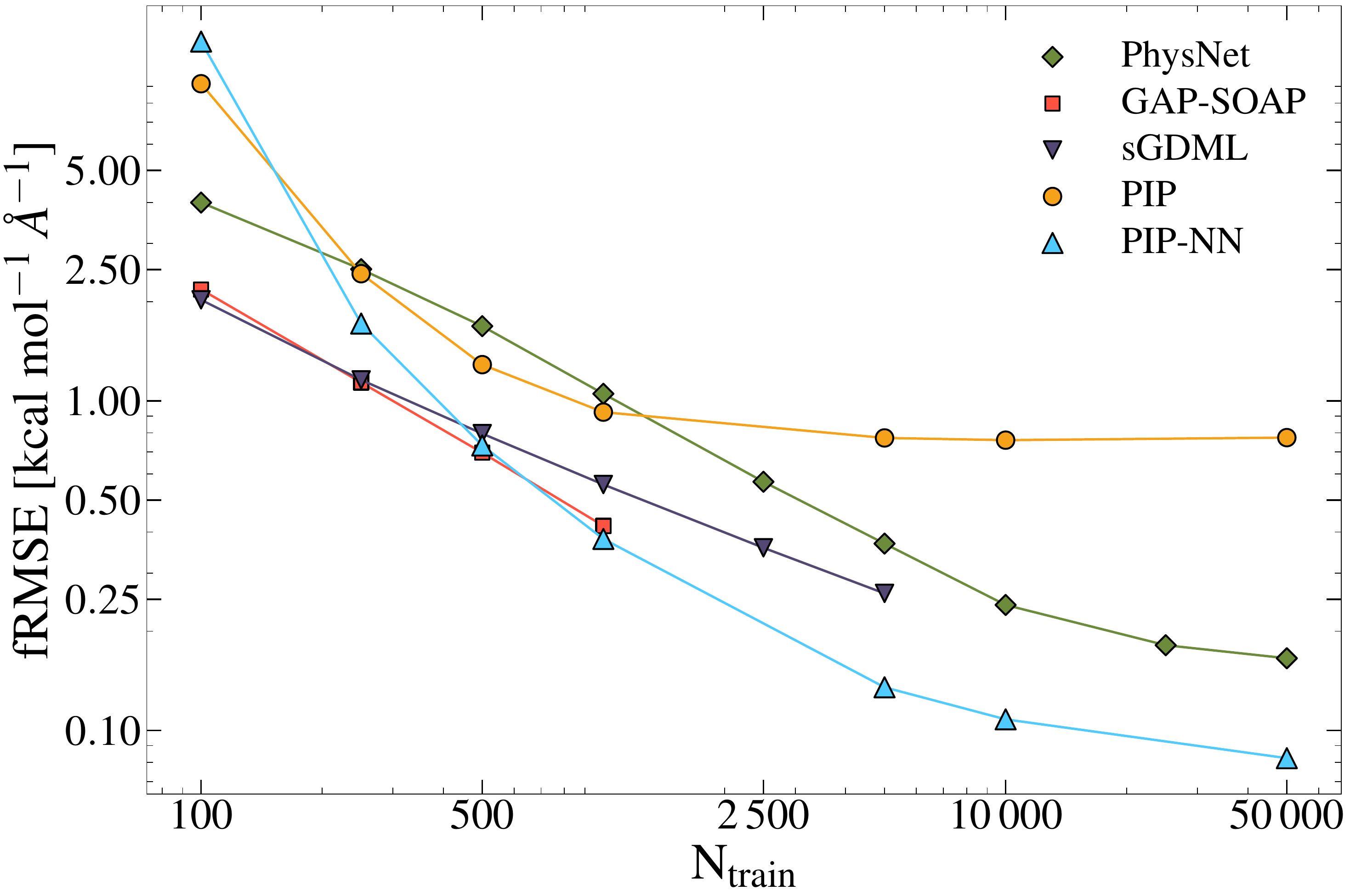}
    \caption{Root-mean-squared errors in energies (left panel) and forces (right panel) against the budget for training and validation for various ML PESs trained on the MD17-ethanol data set. The RMSE values were calculated on a set of 100,000 points separate from the training or validation sets.}
    \label{fig:ethanol}
\end{figure}

We adopted the MD17-ethanol database\footnote{The data set was obtained from \url{http://quantum-machine.org/gdml/#datasets}} to assess the performance of our implementation of the PIP-NN method. It is a formidable task to properly compare the performances of ML PESs in terms of accuracy and computational efficiency. To guide our benchmark study in terms of accuracy, we address the protocol suggested by \citet{Pinheiro2021} based on the analysis of learning curves.
With the exception of the sGDML approach, which can be fit to gradients only, we consider the accuracy of several models trained on data sets of increasing size that contain both energies and forces. Figure~\ref{fig:ethanol} compares the root-mean-squared errors for the energies (eRMSE) and forces (fRMSE) evaluated on a set-aside test set as a function of the training and validation budget. 
To construct a full PIP basis, we employ the permutational symmetry group 321111 with an overall polynomial order of 3, which yields a basis of size 1898. The basis set was used as is; no truncations outlined in the Methods section were applied.
First, we build a model (referred to as ``PIP" in Figure~\ref{fig:ethanol}) as a linear combination of PIPs with coefficients tuned to minimize the loss function~\eqref{methods:loss-forces}. Note that the PIP-NN method formally becomes a PIP regression if the neural network is reduced to a single output neuron with a linear transfer function.
Also trained with the loss function~\eqref{methods:loss-forces}, the PIP-NN model uses a neural network consisting of a single hidden layer with 32 nodes -- an architecture we denote as (1898-32-1) -- which integrates nearly 60\,000 parameters. The performance values for the PhysNet, GAP-SOAP, and sGDML models are taken from the \citet{Pinheiro2021} study. For large enough training sets, all approaches can achieve RMSE values well below 1 kcal/mol, which is usually referred to as "chemical accuracy". Achieving an eRMSE and a fRMSE of less than 0.05 kcal/mol and 0.1 kcal/mol/\AA, respectively, the PIP-NN models manage to attain the best accuracy in energies and forces over the examined range of training set sizes among the considered models. Also, for the same PIP basis, at sufficiently high training sizes, the PIP-NN attains close to 5 and 10 times smaller eRMSE and fRMSE values, respectively, compared to PIP regression model. This highlights the benefits of building a fully-connected neural network on top of the PIP basis.

Given the plethora of various technical details related to the implementation of ML PESs, it is rather challenging to assess their computational efficiency on an equal footing. Here, we forego conducting comprehensive study and limit our focus to demonstrating the effectiveness of our implementation. Building on the assessment made by \citet{Houston2022}, we consider the time needed to predict energies and gradients for a set of 20\,000 configurations.
It is worth reminding that memory requirements and prediction times for ML PESs based on kernel methods, such as sGDML and GAP-SOAP, grow with increasing training sets. \citet{Houston2022} reports the prediction timings of 1\,100 and 195 seconds for GAP-SOAP and sGDML methods for the training set sizes of 1\,000 and 5\,000, respectively. The prediction cost for parametric approaches such as PhysNet and PIP depends on the chosen architecture or basis set rather than on the amount of data. PhysNet and the PIP model (based on the basis of 1898 invariant polynomials) give the predictions for energies and forces for a reference set in 214 and 0.23 seconds, respectively. All of the mentioned results were obtained by \citet{Houston2022} on a system with an Intel Xeon Gold 6240 processor. We ran our benchmark on an Intel Xeon Gold 6148, which has a slightly higher number of cores (20 vs. 18) with somewhat lower single-core performance, resulting in similar multi-core performance. The set of targets is predicted by the PIP-NN model with (1898, 32, 1) architecture in 0.35s. Performance-wise, we observe that PIP-NN is within the same order of magnitude as the PIP model and around two orders of magnitude more robust than the quickest alternative, even though the prediction times obtained on different machines could only be tentatively compared.

\section{Summary and Conclusions}
\label{sec:conclusions}

In this study, we develop the machine-learned two-body PESs of non-covalent interactions within N$_2-$Ar and N$_2-$CH$_4$ systems with flexible monomers. To this end, we make use of the PIP-NN approach proposed by Guo and co-workers \cite{Jiang2013}, in which a collection of permutationally invariant polynomials is employed as the input of the fully-connected neural network. To take into account the fundamental characteristics of the two-body term, we explored several modifications to the model framework by specifically tailoring a set of PIPs.
Starting from the full basis set of invariant polynomials transforming according to the given permutational symmetry group, the purified basis is obtained by eliminating the polynomials that do not able to ensure the asymptotic zero-interaction limit of the two-body term. The pruned purified basis is obtained upon further reducing the previously purified basis to those polynomials that depend exclusively on the intermonomer distances. The PIP-NN models built on top of the purified and pruned purified basis sets exhibit correct asymptotic behavior. We also address the behavior of the PIP-NN model as it approaches the zero-interaction limit. To improve the accuracy of the model in the long range region, we employ a mixed Morse-polynomial form of the bond-order variables. 
The PP basis set naturally emerges as a suitable tool for constructing a model of two-body PES within rigid-rotor approximation. We demonstrate that the PIP-NN N$_2-$CH$_4$ model, which is predicated on the PP basis set, could exceed in accuracy the regression model based on the symmetry-adapted angular functions that we reported in previous effort \cite{Finenko2021}.
The parameters entering the PIP-NN models for N$_2-$Ar and N$_2-$CH$_4$ were determined by training on the data sets of interaction energies calculated at CCSD(T)/CBS and CCSD(T)-F12a levels of theory, respectively. 
The second virial coefficient calculated as a function of temperature within classical framework for both N$_2-$Ar and N$_2-$CH$_4$ is found in good agreement with the available experimental data. With eRMSE values of 0.35 and 0.85 cm$^{-1}$ for interaction energies up to 40,000 cm$^{-1}$ and 10,000 cm$^{-1}$, respectively, the constructed PESs are of high fidelity.
For the sake of comparison, a PIP regression approach is reported to attain an eRMSE of 3.5 cm-1 for the H$_2$O$-$CH$_4$ complex \cite{Qu2015}, which is similar in dimensionality to N$_2-$CH$_4$. To reach this accuracy, however, an extensive PIP basis with over 10\,000 polynomials was used. In turn, the 524 polynomials of the purified basis set -- which is thus about twenty times smaller -- were employed in full-dimensional PIP-NN model for N$_2-$CH$_4$.
The high level of fidelity achieved by the PIP-NN model is due to the large number of parameters packed into the neural network. Relative to the computational cost associated with evaluating PIPs of a large basis set, it is important to emphasize that processing time of feed-forwarding the values across the network is rather low. Accordingly, from our perspective, one of the primary benefits of the PIP-NN approach is that it enables the construction of high-fidelity PESs from more compact collections of invariant polynomials than the regression method.

To further this claim we compare the PIP-NN approach against the popular state-of-the-art ML PESs, namely, sGDML, PhysNet, GAP-SOAP and PIP regression model, on the MD17 database of energies and forces for ethanol. 
Recent thorough assessment of these and several others methods were conducted by \citet{Pinheiro2021} for the MD17-ethanol database. We adopted the protocol based on the analysis of learning curves suggested in the assessment study.
The PIP-NN model with (1898, 32, 1) architecture achieves an eRMSE an a fRMSE of less than 0.05~kcal/mol and 0.1~kcal/mol/\AA, respectively, and so attains the best accuracy among the considered models using the budget for training and validation consisting of 50\,000 configurations. We also show that  the PIP-NN model's eRMSE and fRMSE values are almost 5 and 10 times lower than those obtained by the PIP regression model, which is based on the same set of invariant polynomials. 
From the performance standpoint, our implementation of the PIP-NN model runs roughly 50\% slower than the \citet{Houston2022}'s implementation of the PIP model (accounting for small differences in Intel CPUs used). The performance was evaluated by predicting energies and forces for a collection of 20\,000 target configurations. Other models have substantially higher prediction times, especially when larger training and validation budgets are considered. Ultimately, the most robust non-PIP model, sGDML, runs almost three orders of magnitude slower than the PIP-NN and PIP regression models.

\section{Data sets and code availability}
\label{sec:dataset-and-code}

The interaction energy data sets for N$_2-$Ar and N$_2-$CH$_4$ as well as code for training the PIP-NN models is freely available at \url{https://github.com/artfin/PES-Fitting-MSA}.

\section{Acknowledgements}
\label{sec:acknowledgements}

The research was supported by the non-commercial Foundation for the Advancement of Science and Education INTELLECT. I acknowledge the support by Grant No. RSCF 22-17-00041 in developing machine-learned two-body PES for the N$_2-$CH$_4$ system.
We were able to carry out \textit{ab initio} calculations and train ML PESs thanks to the Smithsonian Institution's allocation of computing time on the HPC and GPU facilities of the Smithsonian Institution High Performance Cluster\cite{SIHPC}. Additionally, I am thankful to Daniil Chistikov and Andrey Vigasin for their insightful comments and helpful discussions.

\bibliography{biblio} 

\end{document}